\documentclass{mem}

\usepackage{natbib}
\usepackage{txfonts}
\usepackage{balance}
\usepackage{graphicx}
\usepackage{epstopdf}
\usepackage{epsfig}
\usepackage[a4paper,breaklinks]{hyperref}

\begin{document}
  
  \def\teff{$T\rm_{eff }$}
  \def\kms{$\mathrm {km s}^{-1}$}

  \title{
	  The impact of M-dwarf atmosphere modelling\\ on planet detection
	}
	
  \subtitle{}

    \author{
	      I. \,Bozhinova
	      \and Ch. \,Helling
	      \and C. \,Stark
          }

    \offprints{Ch. Helling, ch@leap2010.eu}

    \institute{
		SUPA, School of Physics and Astronomy, University of St Andrews, North Haugh, St Andrews,  KY16 9SS, UK		
	      }


  
  \abstract{
    Being able to accurately estimate stellar parameters based on spectral observations is important
    not only for understanding the stars themselves but it is also vital for the determination of exoplanet parameters. 
    M dwarfs are discussed as targets for planet detection as these stars are less massive, 
    less luminous and have smaller radii making it possible to detect smaller and lighter planets.  
    Therefore M-dwarfs could prove to be a valuable source for examining the lower mass end of planet distribution,
    but in order to do that, one must first take care to understand the characteristics of the host stars well enough.
    
    Up to date, there are several families of stellar model atmospheres. 
    We focus 
    on the ATLAS9, MARCS and {\sc Drift-Phoenix} families in the M-dwarf parameter space.
    We examine the differences in the (T$_{\rm gas}$, p$_{\rm gas}$)  structures, synthetic photometric fluxes and related colour indices.
    We find discrepancies in the hotter regions of the stellar atmosphere between the ATLAS and MARCS models. 
    The MARCS and {\sc Drift-Phoenix} models appear to agree to a better extend with variances of less than 300K. 
    We have compiled the broad-band synthetic photometric fluxes of all models for the Johnson UBVRI and 2MASS JHKs.
    The fluxes of MARCS differ from both ATLAS and {\sc Drift-Phoenix} models in the optical range.

  }
  
 \maketitle{}

 \section{Introduction}
    The hunt for exoplanets has been on for about 20 years since the discovery of the very first exoplanets (\citealt{wol}, \citealt{mayor}, \citealt{charb}). 
    Up to date, there are 872 confirmed planets discovered in 683 planetary systems ({\it exoplanet.eu}). 
    With various surveys, involving high-precision instruments, such as {\sc HARPS} (\citealt{mayor2}), {\sc COROT} (\citealt{auvergne}) or {\sc KEPLER} (\citealt{batalha}) it 
    is no surprise that, in the last two years, astronomers have detected more and more planets within the so-called Super-Earth group.
    However, our knowledge of a given planetary system dependents on the knowledge of its host star. 
    This study is dedicated to the modeling of M-dwarf atmospheres and the implications these models could pose in relation to exoplanets.

 \section{Models used}
  The choice of models for this work consists of the ATLAS9 models (\citealt{kurucz}, \citealt{castelli}), the MARCS models (\citealt{gustafsson}) 
  and the {\sc Drift-Phoenix} models (\citealt{dehn}, \citealt{hellingb}, \citealt{witte}). 
  All of these models obey LTE, hydrostatic and chemical equilibrium and energy flux conservation; they are homogeneous, 1D codes that assume plane-parallel symmetry. 
  
  The 
  ATLAS models used here span the range for log(g) = 3.0$\,\ldots\,$5.0, [Fe/H] = $+0.5\,\ldots\, -2.5$
  and effective temperature T$_{\rm eff}$ = 3500 K$\,\ldots\,$4000 K. All these models are calculated with the convection option switched on but with the overshooting option switched off. A mixing length height with l/H$_p$= 1.25 and v$_{\rm turb}$ = 2.0 km/s is adopted for all models.
 
  The MARCS models used span log(g) = 3.0$\,\ldots\,$5.0, [Fe/H] = $+0.5\,\ldots\,-2.5$, and T$_{\rm eff}$=3500K and 4000K in order to allow a  comparison with the ATLAS models. Other MARCS models used are
  T$_{\rm eff}$ =  2500K$\,\ldots\,$3000K, log(g) = 3.0$\,\ldots\,$5.5, and [Fe/H]=0.0 for comparison with the {\sc Drift-Phoenix} models.  For all models, v$_{\rm turb}$= 2 km/s and 
  solar element abundances (\citealt{Grevesse})  were chosen.
  
  The {\sc Drift-Phoenix} models are aimed specifically at late-type stars (M-dwarfs, brown dwarfs) and giant planet atmospheres as they  include a model of dust cloud formation.
  The subset of models used is for the solar metalicity models with 2500K  $<$  T$_{\rm eff} <$  3000K and 3.0 $<$ log(g) $<$ 5.5 .
  
  The different model families  have a different coverage of the M-dwarf regime.
 We therefore have analysed pairs of models, in particular the ATLAS$+$MARCS models for T$_{\rm eff}$ = 3500K and T$_{\rm eff}$ = 4000K and varying log(g) and [Fe/H] values,
  as well as the MARCS+Phoenix models for solar metalicity and varying T$_{\rm eff}$ and log(g) values. A total of 105 models were considered.
 
  \begin{figure*}[]
    \resizebox{\hsize}{!}{\includegraphics[clip = true, angle = 180]{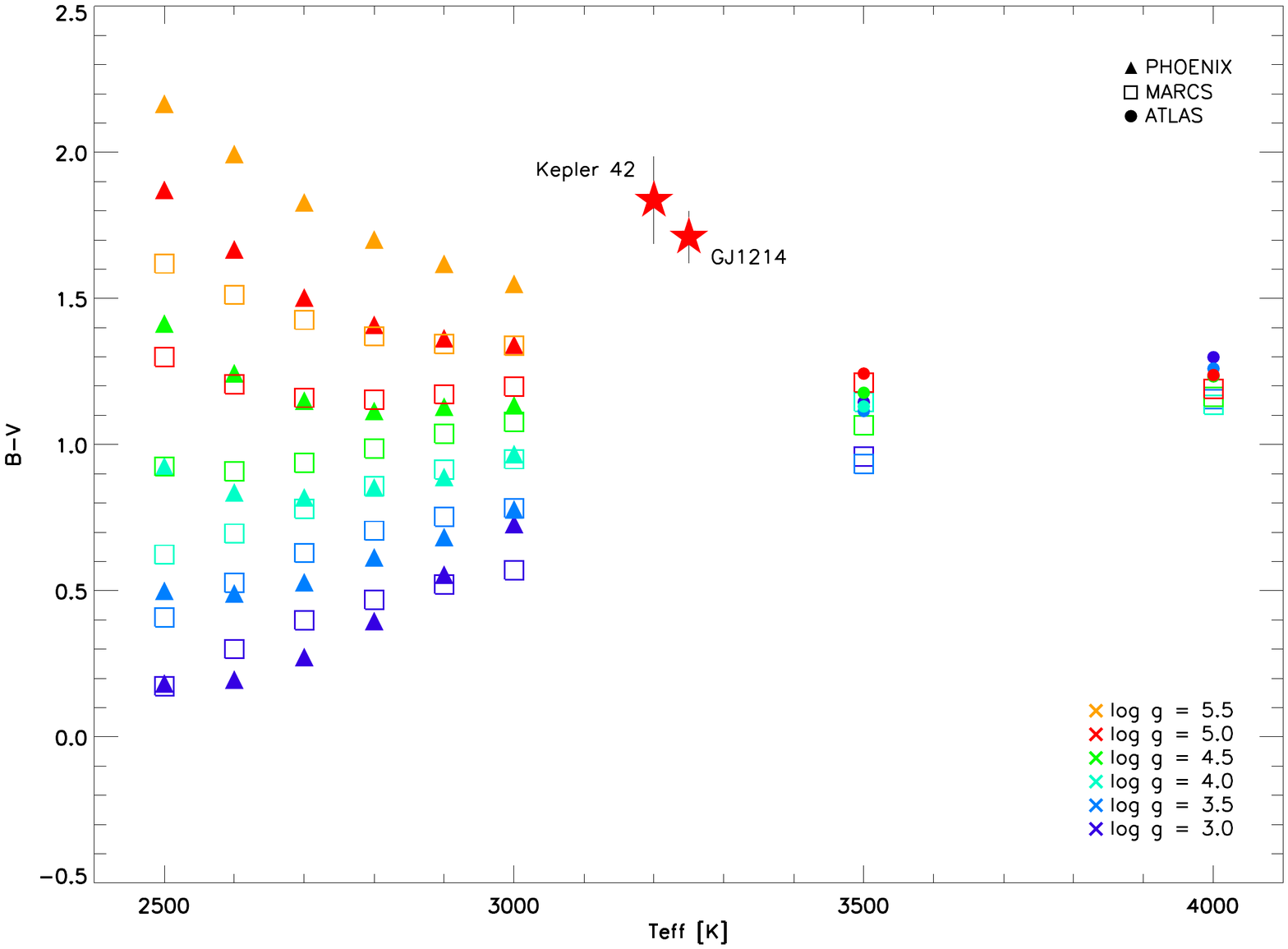}}
    \caption{
	      \footnotesize
	      B-V versus effective temperature for all model families (circles - ATLAS, squares - MARCS, triangles - {\sc Drift-Phoenix}) of solar metalicity.  
	      Colour coding is used for different values for log(g) with a step of 0.5 dex. 
	      The two red stars represent observed data for GJ 1214 and Kepler 42 (\citealt{GJ1214}; \citealt{Kepler42}).
	      }
    \label{B-V}
  \end{figure*}

 \section{Atmospheric structure comparison}
  A comparison of the atmospheric structures of the different models 
   includes 
   the (T$_{\rm gas}$, p$_{\rm gas}$) structures of model atmospheres with matching T$_{\rm eff}$, log(g) and [Fe/H] values,  
  and the differences in the opacity structures.
  
  We look at the residuals in local temperature versus local pressure between the ATLAS and MARCS models for T$_{\rm eff}$ = 3500K and T$_{\rm eff}$ = 4000K. 
  It is interesting to note that while the 3500K models seem to match better in the metalicity range -1.5 $<$ [Fe/H] $<$ -2.5, the 4000K models display better agreement for [Fe/H]=+0.5 and [Fe/H]=0.0. 
  For both T$_{\rm eff}$, the biggest discrepancies lie within the [Fe/H]=-1.0 models, with differences reaching over 1500K in the T$_{\rm eff}$=3500K and over 1200K for the T$_{\rm eff}$=4000K. 
  We create residual plots for each pair of matching models by subtracting local temperature values at matching local pressure values by interpolating models where neccessary. In all cases, the models appear to diverge as the pressure increases,
  i.e. as going deeper into the atmosphere, regardless of particular values for T$_{\rm eff}$, log(g) or [Fe/H]. These trends are also reflected in the Rosseland mean opacities,
where higher divergence in opacity residuals is observed for the same model parameter values described in this paragraph.
  
  The MARCS and {\sc Drift-Phoenix} grids have common models for only one metalicity ([Fe/H]=0.0) but for various effective temperatures (2500-3000K). 
  We observe better agreement between MACRS and PHOENIX than there is between the MACRS and ATLAS models with respect to residual values.
  Overall, this set of models do not appear to vary by more than 300K (with the exception of the case for T$_{\rm eff}$=3000K and log(g)=3.0 and 3.5 ). 
  That is again confirmed by the Rosseland mean opacity residuals. 
  
  In summary, we find that for the higher effective temperature values (3500K, 4000K) the ATLAS and MARCS temperature-pressure structures diverge from each other with an average of $\sim$600K in local temperature
  and extreme cases well over 1000K. In contrast, the MARCS and {\sc Drift-Phoenix} only differ by $\sim$300K for for 2500K $<$ T$_{\rm eff}$ $<$ 3000K. A direct comparison between
  ATLAS and {\sc Drift-Phoenix} is currently not possible as these models do not share any common T$_{\rm eff}$ values.

 \section{Synthetic photometry}
  The (T$_{\rm gas}$, p$_{\rm gas}$)  structure determines the emergent spectral energy distribution for stars.
  In order to compare the SEDs of the different models, we perform synthetic photometry of all models considered. We convolve  the model SEDs to the Johnson UBVRI and 2MASS JHKs filter systems, using HST spectrum of Vega (\citealt{vega}) for zero-point calibration.
  
  We compare the ratios between the synthetic broad-band fluxes for all pairs of corresponding models in each filter. 
  A value of 1.0 corresponds to perfect match.
  The broad-band fluxes of ATLAS and MARCS in the optical (Johnson UBVR filters) differ significantly more than those in the IR range.
  The flux ratios for T$_{\rm eff}$ = 3500K are deviating from 1.0 significantly more (as high as $\sim$1.8)
  than those for T$_{\rm eff}$=4000K (less than $\sim$1.3). There is also a discrepancy between the {\sc Drift-Phoenix} and MARCS models at the optical wavelengths.
  What is more, the spread in ratios appears to be even bigger, with the highest values almost reaching a factor of 2.0 . 
  Unfortunately, the ATLAS and {\sc Drift-Phoenix} families do not have any models in common to allow for a direct comparison.
  
  Finally we compile a list of the synthetic colour indices for each model, using the calculated visual magnitudes (Fig.\ref{B-V}).
   The B-V magnitudes differ by up to half a magnitude between the {\sc Drift-Phoenix} and MARCS models in the low temperature half of the plot 
  and a trend is difficult to place on the data as the curves appear to be very different for these two models. 
  The difference seems to diminish, however, when the models move to higher  T$_{\rm eff}$ as the MARCS and ATLAS models appear to diverge much less, especially at T$_{\rm eff}$=4000K. 
 The two available data points from actual observations, Kepler 42 (Muirhead et al. 2012) and GJ1214 (Anglada-Escude et al. 2013) 
  do not appear to lie on any of the theoretical curves we have compiled, which calls for an extension of the study presented here.

 \section{Conclusion}
  As the field of exoplanetary detection drifts towards Earth-sized planets, the observational community will benefit from targeting M-dwarfs as host stars. 
  This study has compared the temperature-pressure and opacity structures of three model atmosphere families.
  The ATLAS and MARCS models have shown increasing discrepancies in both local temperature and opacity as one goes deeper into the stellar atmosphere. 
  The MARCS and {\sc Drift-Phoenix} models have shown better agreement with local temperature differences of no more than 300K. 
  The MARCS models display considerable deviations from both ATLAS and {\sc Drift-Phoenix} in the optical regime in terms of sythetic photometric fluxes, which is further confirmed in the B-V plots. 
  Even so, there are big gaps in the availability of models, which need to be filled in order to present a more comprehensive study.

 \bibliographystyle{aa}

\end{document}